\documentclass[12pt]{article}
\usepackage{amsfonts}

\usepackage[dvips]{graphicx}
\usepackage{epsfig}
\usepackage{amsmath}
\usepackage{float}

\newcommand{\beqs}{\begin{eqnarray}}
\newcommand{\eeqs}{\end{eqnarray}}

\newcommand{\ra}{\rightarrow}

\begin{document}

\thispagestyle{empty}

\begin{flushright}
hep-th/0408234
\end{flushright}

\hfill{}

\hfill{}

\hfill{}

%\vspace{32pt}

\begin{center}
\textbf{\Large On the Thermodynamics of Nut Charged Spaces } \\[0pt]

\vspace{20pt}
{\bf Robert B. Mann}\footnote{E-mail: 
{\tt mann@avatar.uwaterloo.ca}}$^{,2}$
{\bf and Cristian Stelea}\footnote{E-mail: 
{\tt cistelea@uwaterloo.ca}}

\vspace*{0.2cm}

{\it $^{1}$Perimeter Institute for Theoretical Physics}\\
{\it 31 Caroline St. N. Waterloo, Ontario N2L 2Y5 , Canada}\\[.5em]

{\it $^{2}$Department of Physics, University of Waterloo}\\
{\it 200 University Avenue West, Waterloo, Ontario N2L 3G1, Canada}\\[.5em]
\end{center}

\vspace{20pt}

\begin{abstract}
\addtolength{\baselineskip}{1.2mm} We discuss and compare at length the
results of two methods used recently to describe the thermodynamics of
Taub-NUT solutions in a deSitter background. In the first approach ($\mathbb{%
C}$-approach), one deals with an analytically continued version of the metric
while in the second approach ($\mathbb{R}$-approach), the discussion is
carried out using the unmodified metric with Lorentzian signature. No
analytic continuation is performed on the coordinates and/or the parameters
that appear in the metric. We find that the results of both these approaches
are completely equivalent modulo analytic continuation and we provide the
exact prescription that relates the results in both methods. The extension
of these results to the AdS/flat cases aims to give a physical interpretation of the thermodynamics of nut-charged spacetimes in the Lorentzian sector. We also briefly discuss the higher dimensional spaces and note that, analogous with the absence of hyperbolic nuts in AdS backgrounds, there are no spherical Taub-Nut-dS solutions.
\end{abstract}

\setcounter{footnote}{0}

\newpage

\section{Introduction}

The construction of conserved charges in a deSitter (dS) background is in
general not well-defined for several reasons. One is that such spaces do not
have spatial infinity the way that asymptotically flat or anti de Sitter
(AdS) spacetimes do. Another reason is that it is impossible to define a global
Killing vector that is everywhere timelike. In fact, there is a Killing
vector that is timelike inside the cosmological horizon and spacelike
outside it.

However, motivated in part by the possible analogy with the AdS/CFT
correspondence, we expect to be able to construct holographic duals of
asymptotically dS spacetimes \cite{strominger}. The specific prescription
given in ref. \cite{Balasubramanian:2002wy} introduced counterterms on spatial
boundaries at past and future infinity that yielded a finite action for
deSitter backgrounds. One can thus compute a stress tensor on the
past/future boundary and consequently a conserved quantity associated with
each Killing vector there.

The conserved charge associated with the Killing vector $\frac{\partial }{\partial t}$ is interpreted as the conserved mass. With this definition, the maximal mass conjecture -- that any asymptotically dS spacetime with mass greater than that of pure dS space has a cosmological singularity -- was
proposed in ref. \cite{Balasubramanian:2002wy}. This conjecture was partly based on
the Bousso N-bound \cite{Bousso1,Bousso2}, which is the conjecture that any
asymptotically dS spacetime will have an entropy no greater than the entropy 
$\pi l^{2}$ of pure dS with cosmological constant $\Lambda =\frac{3}{l^{2}}$.

The maximal mass conjecture has been verified for topological dS solutions
and their dilatonic variants and for the Schwarzschild-dS black hole up to
nine dimensions \cite{GM,Cai,Astefanesei} and, recently, for the class of
Kerr-deSitter spacetimes \cite{GM2}; however, it is violated (along with the N-bound
conjecture) in locally asymptotic deSitter spacetimes with NUT charge under
certain conditions \cite{rick2, rick1,Anderson}.

The preceding results rely on the relationships between various
thermodynamic quantities at past/future infinity. In asymptotically flat
or AdS settings, such relationships may be established using the
path-integral formalism of semi-classical quantum gravity \cite{flatAdSrefs}. For spaces with rotation or NUT charge, such relationships depend upon how one analytically continues these parameters into a Euclidean section.

In asymptotically dS spacetimes the situation is even more delicate.
Analytic continuation in the rotating case requires special care \cite{boothmart}. More recently it was noted that although the computation of conserved quantities does not depend upon such analytic continuation, the path-integral foundations of thermodynamics at asymptotic past/future
infinity does, and that there are two apparently distinct ways of expressing
the metric, depending on which set of Wick rotations is chosen \cite{rick2,rick1}.

In one approach (referred to as the $\mathbb{R}$-approach), the analysis is
carried out using the unmodified metric with Lorentzian signature; no
analytic continuation is performed on the coordinates and/or the parameters
that appear in the metric.

In the alternative $\mathbb{C}$-approach one deals with an analytically
continued version of the metric and at the end of the computation all the final results are analytically continued back to the Lorentzian sector. However, the simple analytic continuation of
Euclidian time into the Lorentzian time $T\rightarrow it$ fails in many physically interesting
cases (for example, if the spacetimes are manifestly not static, or even in the stationary cases). While this is a known issue in rotating spacetimes \cite{boothmart}, it is shown most strikingly in the case of nut-charged spacetimes. Consider for example the asymptotically flat Taub-NUT spacetime. As it is well known \cite{misner,Page,mann1}, this spacetime is nonsingular only if we make the time coordinate $t$ periodic with period $8\pi n$, in order to eliminate the Misner string singularity. To obtain the Euclidian
sector we perform the analytic continuation of the time coordinate $t\rightarrow iT$ and of the nut parameter $n\rightarrow iN$. To keep the Euclidian section non-singular, that is in order to eliminate a possible conical singularity that would appear in the $\left( r,T\right) $ section, a
constraint relating the mass to the nut charge must be imposed, consistent
with the preceding periodicity requirements (which imply $\beta =8\pi N$).
For the Taub-Nut solution this is $m=N$ whereas for the Taub-Bolt solution $m=\frac{5}{4}N$ \cite{Page}. However, the physical interpretation of the results in terms of the parameters appearing in the Lorentzian sector is somewhat problematic since a naive analytic continuation would send $T\rightarrow it$ and $N\rightarrow in$ and render imaginary the physical quantities of interest.

Nut-charged spacetimes do have the reputation of being a `counterexample to
almost anything' in General Relativity \cite{misner}, so that it is natural to
regard the above analytic continuations with suspicion. Certainly the
presence of the closed timelike curves in the Lorentzian sector is a less
than desirable feature to have. However, it is precisely this feature that
makes them more interesting. Recently these spaces have been receiving increased
attention as testbeds in the AdS/CFT conjecture \cite{Hawking2,myers1}.
Another very interesting result concerns a non-trivial embedding of the
Taub-NUT geometry in heterotic string theory, with a full conformal field
theory definition (CFT) \cite{johnson}. It was found that the nutty effects
were still present even in the exact geometry, computed by including all the
effects of the infinite tower of massive string states that propagate in it.
This might be a sign that string theory can very well live even in the
presence of nonzero nut charge, and that the possibility of having closed
timelike curves in the background can still be an acceptable physical
situation. Furthermore, in asymptotically dS settings, regions near
past/future infinity do not have CTCs, and nut-charged asymptotically dS
spacetimes have been shown to yield counter-examples to some of the
conjectures advanced in the still elusive dS/CFT paradigm \cite{strominger}-
such as the maximal mass conjecture and Bousso's entropic N-bound conjecture
(for a review see \cite{rickreview}). For these reasons we believe that a
more detailed study of these spaces and, in particular, a study of the
thermodynamics of nut-charged spacetimes is worthwhile and still has new
things to teach us, if only to point out where our theories can fail.

Our main goal in this paper is to clarify the relationship between the $\mathbb{R}$ and $\mathbb{C}$ approaches, with an eye toward understanding how to physically interpret the results obtained in each case. We find that the results of both these approaches are completely equivalent modulo analytic continuation. Furthermore, we provide an exact prescription that relates the results in both methods. Extending our methods to asympotically AdS/flat cases yields a physical interpretation of the thermodynamics of nut-charged spacetimes in the Lorentzian sector. We discuss the constraints that appear by imposing the first law of thermodynamics. We find that the first law will hold precisely for the (asymptotically AdS) Bolt and Nut solutions that we obtain in the $\mathbb{C}$-and $\mathbb{R}$-approaches, which we take as a sign of the validity of our results regarding the thermodynamics of the Lorentzian Taub-NUT-(A)dS solutions. We also briefly discuss the case of higher dimensional nut-charged spacetimes.

This paper is organized as follows. Since our approach to the thermodynamics of the nut-charged spaces is based on the direct application of the Gibbs-Duhem relation, in the next section we recall the counterterm method, which is used to compute the conserved quantities in asymptotically dS-spacetimes. Next we briefly review the thermodynamics of $4$-dimensional Taub-NUT-dS, using both the $\mathbb{C}$-and $\mathbb{R}$-approaches. Upon comparison we find complete equivalence between these approaches modulo an analytic continuation and we provide the prescription that relates the results in both cases. This prescription of analytic continuation furnishes the connection between the thermodynamics in Lorentzian signature ($\mathbb{R}$-approach) and the thermodynamics of the Taub-NUT solutions in the analytically continued regime ($\mathbb{C}$-approach) in which the signature is no longer Lorentzian. We consider the extension of our results to higher dimensional nut-charged spaces as well as to the case of the Taub-NUT solutions in $AdS$ and flat backgrounds. The final section is dedicated to some concluding remarks.

\section{The counterterm method in deSitter backgrounds}

For a general asymptotically dS spacetime, the action can be decomposed into
three distinct parts 
\begin{equation}
I=I_{B}+I_{\partial B}+I_{ct}  \label{actiongeneral}
\end{equation}%
where the bulk ($I_{B}$) and boundary ($I_{\partial B}$) terms are the usual
ones, given by 
\begin{eqnarray}
I_{B} &=&\frac{1}{16\pi }\int_{\mathcal{M}}d^{d+1}x~\sqrt{-g}\left({R}%
-2\Lambda +\mathcal{L}_{M}(\Phi )\right)  \label{actionbulk} \\
I_{\partial B} &=&-\frac{1}{8\pi }\int_{\partial \mathcal{M}^{\pm }}d^{d}x~%
\sqrt{\gamma ^{\pm }}\Theta ^{\pm }  \label{actionboundary}
\end{eqnarray}%
where $\partial \mathcal{M}^{\pm }$ represents future/past infinity, and $%
\int_{\partial \mathcal{M}^{\pm }}=\int_{\partial \mathcal{M}^{-}}^{\partial 
\mathcal{M}^{+}}$ represents an integral over a future boundary minus an
integral over a past boundary, with the respective metrics $\gamma ^{\pm }$
and extrinsic curvatures $\Theta ^{\pm }$ (working in units where $G=1$).
The quantity $\mathcal{L}_{M}(\Phi )$ in (\ref{actionbulk}) is the
Lagrangian for the matter fields, which we shall not be considering here.
The bulk action is over the $(d+1)$-dimensional manifold $\mathcal{M}$, and the
boundary action is the surface term necessary to ensure well-defined
Euler-Lagrange equations. For an asymptotically dS spacetime, the boundary $%
\partial \mathcal{M}$ will be a union of Euclidean spatial boundaries at
early and late times.

The counter-term action $I_{ct}$ in (\ref{actiongeneral}) appears in the
context of the dS/CFT correspondence conjecture due to the counterterm
contributions from the boundary quantum CFT \cite{balakraus,CFTref}. It has
a universal form for both the AdS and dS cases and it can be generated by an
algorithmic procedure, without reference to a background metric, with the
result \cite{GM} 
\begin{eqnarray}
I_{ct} &=&-\int d^{d}x\sqrt{\gamma }\bigg[-\frac{d-1}{l}+\frac{l\mathsf{%
\Theta }(d-3)}{2(d-2)}\mathsf{R}-\frac{l^{3}\mathsf{\Theta }(d-5)}{%
2(d-2)^{2}(d-4)}\left( \mathsf{R}_{ab}\mathsf{R}^{ab}-\frac{d}{4(d-1)}%
\mathsf{R}^{2}\right)  \notag \\
&&-\frac{l^{5}\mathsf{\Theta }(d-7)}{(d-2)^{3}(d-4)(d-6)}\left( \frac{3d+2}{%
4(d-1)}\mathsf{RR}^{ab}\mathsf{R}_{ab}-\frac{d(d+2)}{16(d-1)^{2}}\mathsf{R}%
^{3}\right.  \notag \\
&&\left. -2\mathsf{R}^{ab}\mathsf{R}^{cd}\mathsf{R}_{acbd}-\frac{d}{4(d-1)}%
\nabla _{a}\mathsf{R}\nabla ^{a}\mathsf{R}+\nabla ^{c}\mathsf{R}^{ab}\nabla
_{c}\mathsf{R}_{ab}\right) +\ldots \bigg]  \label{actionct}
\end{eqnarray}%
with $\mathsf{R}$ the curvature of the induced metric $\gamma $ and $\Lambda
={\textstyle\frac{d(d-1)}{2l^{2}}}$. The step-function $\mathsf{\Theta }%
\left( x\right) $ is unity provided $x>0$ and vanishes otherwise. For
example, in four ($d=3$) dimensions, only the first two terms appear, and
only these are needed to cancel divergent behavior in $I_{B}+I_{\partial B}$
near past and future infinity.

Varying the action with respect to the boundary metric $h_{ij}$ gives us the
boundary stress-energy tensor: 
\begin{eqnarray}
T^{\pm ij}&=&\frac{2}{\sqrt{h^{\pm}}}\frac{\delta I}{\delta h^{\pm ij}}
\end{eqnarray}
If the boundary geometries have an isometry generated by a Killing vector $%
\xi ^{\pm \mu }$, then $T_{ab}^{\pm }\xi^{\pm b}$ is divergence free, from
which it follows that the quantity 
\begin{equation}
\mathfrak{Q}{}^{\pm }=\oint_{\Sigma ^{\pm }}d^{d-1}\varphi ^{\pm }\sqrt{%
\sigma ^{\pm }}n^{\pm a}T_{ab}^{\pm }\xi ^{\pm b}  \label{Qcons}
\end{equation}
is conserved between histories of constant $t$, whose unit normal is given
by $n^{\pm a}$. The $\varphi ^{a}$ are coordinates describing closed
surfaces $\Sigma $, where we write the boundary metric(s) of the spacelike
tube(s) as 
\begin{equation}
h_{ab}^{\pm }d\hat{x}^{\pm a}d\hat{x}^{\pm b}=d\hat{s}^{\pm 2}=N_{T}^{\pm
2}dT^{2}+\sigma _{ab}^{\pm }\left( d\varphi ^{\pm a}+N^{\pm a}dT\right)
\left( d\varphi ^{\pm b}+N^{\pm b}dT\right)  \label{hmetric}
\end{equation}
where $\nabla _{\mu }T$ is a spacelike vector field that is the analytic
continuation of a timelike vector field. Physically this means that a
collection of observers on the hypersurface all observe the same value of $%
\mathfrak{Q}$ provided this surface has an isometry generated by $\xi ^{b}$.

If $\partial /\partial t$ is itself a Killing vector, then we define 
\begin{equation}
\mathfrak{M}{}^{\pm }=\oint_{\Sigma ^{\pm }}d^{d-1}\varphi ^{\pm }\sqrt{%
\sigma ^{\pm }}N_{T}^{\pm }n^{\pm a}n^{\pm b}T_{ab}^{\pm }  \label{Mcons}
\end{equation}
as the conserved mass associated with the future/past surface $\Sigma ^{\pm} 
$\ at any given point $t$ on the spacelike future/past boundary. Since all
asymptotically de Sitter spacetimes must have an asymptotic isometry
generated by $\partial /\partial t$, there is at least the notion of a
conserved total mass ${}\mathfrak{M}^{\pm }$ for the spacetime in the limit
that $\Sigma ^{\pm }$ are future/past infinity.

In order to compute the thermodynamic relationships between conserved
quantities in asymptotically de Sitter spacetimes, we must deal with the
analytic continuation of the metric into a Euclidean section. In this
setting there are several ways to express the metric, depending on which set
of Wick rotations is chosen. In the first approach, one deals with an
analytically continued version of the metric that involves not only a
complex rotation of the (spacelike) $t$ coordinate\thinspace\ ($t\rightarrow
iT$), but also an analytic continuation of the rotation and nut charge
parameters (if any) in the metric, yielding a metric of signature ($%
-,-,+,+,\ldots $). Upon calculation, this will give rise to a negative
action, and hence a negative definite energy. One must also periodically
identify $T$ with period $\beta $ in order to eliminate possible conical
singularities in the $\left( -,-\right) $ section of the metric. This is the
so-called $\mathbb{C}$-approach, since it involves a rotation into the
complex plane. One advantage of using this approach is that $T$ is a `time'
coordinate and the conserved quantity associated with the Killing vector $%
\partial /\partial T$ can be identified unambiguously with a total mass of
the system.

In the $\mathbb{R}$-approach, the analysis is carried out using the
unmodified metric with Lorentzian signature; no analytic continuation is
performed on the coordinates and/or the parameters that appear in the
metric. This option appears because the $t$ coordinate is spacelike outside the cosmological horizon, and so (a semi-classical path-integral) evaluation of thermodynamic quantities at
past/future infinity does not necessarily require its analytical continuation \cite{rick2}. Instead, one evaluates the action at past/future infinity, imposing periodicity in $t$, consistent with regularity at the cosmological horizon (given by the surface gravity of the cosmological horizon of the $\left(
+,-\right) $ section). There is no need to analytically continue either the rotation parameters or nut charges to complex values, and consequently there is no need to analytically continue any results to extract a physical interpretation.

The main thermodynamic relation is an extension of the Gibbs-Duhem relation 
\begin{equation}
S=\beta \mathfrak{M}-I_{cl}  \label{GDuhem}
\end{equation}%
to asymptotically de Sitter spacetimes. For a discussion of its
path-integral foundation, see ref. \cite{rickreview}.

As a simple application of this formalism consider the Scharwzschild-dS
solution in $4$-dimensions, outside the cosmological horizon 
\begin{equation*}
ds^{2}=-\frac{d\tau ^{2}}{F(\tau )}+F(\tau )dt^{2}+\tau ^{2}d\Omega _{2}^{2}
\end{equation*}%
where 
\begin{equation*}
F(\tau )=\left( \frac{\tau ^{2}}{l^{2}}+\frac{2m}{\tau }-1\right) 
\end{equation*}%
Working in the Lorentzian signature (what we call the $\mathbb{R}$-approach
bellow) we obtain the following results for action and the conserved mass %
\cite{GM}: 
\begin{equation*}
I_{SdS}=-\frac{\left( m+\frac{\tau _{+}^{3}}{l^{2}}\right) \beta _{r}}{2},~~~%
\mathfrak{M}=-m,
\end{equation*}%
Here $\tau _{+}$ is the radius of the cosmological horizon and $\beta
_{r}=\int dt$ . 

If the range of the $t$-coordinate is infinite the action will in general
diverge. It is however tempting to impose a periodicity of the time
coordinate even in the Lorentzian sector that is consistent with the
periodicity of the analytically continued time $t\rightarrow iT$.  We
therefore turn to the $\mathbb{C}$-approach, in which the new metric has
signature $(-,-,+,+)$. The sector $(\tau ,T)$ will have a conical
singularity unless the $T$ coordinate is periodically identified with period
\beqs 
\beta _{c}=\frac{4\pi }{|F^{\prime }(\tau _{+})|}
\eeqs
 Since under the analytic continuation $T\rightarrow it$ the periodicity $\beta _{c}$ remains unaffected, there is no obstruction in considering a
similar condition in the Lorentzian sector as well; by continuity we must
require that $\beta _{r}=\beta _{c}$. This will render finite all the
physical quantities of interest and allow a definition of the entropy in the
Lorentzian sector by means of the extended Gibbs-Duhem relation (\ref{GDuhem}%
). The result is $S=\pi \tau _{+}^{2}$, equal to one quarter of the area
of the cosmological horizon.

While the Schwarzschild-dS case is somehow trivial, in the sense that the
equivalence between the  $\mathbb{C}$- and $\mathbb{R}$- approaches fixes
an otherwise arbitrary periodicity in the spacelike $t$ coordinate, this method has been recently extended to the non-trivial case of four-dimensional Kerr-dS spacetimes \cite{GM2}. In nut-charged spacetimes the situation is considerably less trivial, since there are independent
geometric reasons for fixing the periodicity of $t$\ in the $\mathbb{R}$-approach, \textit{i.e.} in the Lorentzian sector. We turn to this situation for the remainder of the paper.

\section{Thermodynamics of Taub-NUT-dS spaces}

In asymptotically (A)dS/flat spacetimes with nut charge there is an additional periodicity constraint for $t$ that arises from demanding the absence of Misner-string singularities. When matched with the periodicity $\beta$, this yields an additional consistency criterion that relates the mass and nut parameters, the solutions of which produce generalizations of asymptotically flat Taub-Bolt/Nut space to the asymptotically (A)dS case. These solutions can be classified by the dimensionality of the fixed point sets of the Killing vector $\xi =\partial /\partial t$\ that generates a $U(1)$ isometry group. In $(d+1)$-dimensions, if this fixed point set dimension is $\left(d-1\right) $ then the solution is called a Bolt solution; if the dimensionality is less than this then the solution is called a Nut solution. If $d=3$,
Bolts have dimension $2$ and Nuts have dimension $0$. However if $d>3$ then Nuts
with larger dimensionality can exist \cite{mann,Lu:2004ya,TNletter}. Note that fixed point sets need not exist; indeed there are parameter ranges of nut-charged asymptotically dS spacetimes that have no Bolts \cite{Anderson}.

It is well known that the Taub-NUT spaces are plagued by quasiregular
singularities \cite{Konk}, which correspond to the end-points of incomplete
and inextensible geodesics that spiral infinitely around a topologically
closed spatial dimension. However, since the Riemann tensor and all its
derivatives remain finite in all parallelly propagated orthonormal frames we
take the point of view that these mildest of singularities are not what is
meant by a cosmological singularity. We shall ignore them when discussing
the singularity structure of the Taub-NUT solutions. We also note that for
asymptotically dS spacetimes that have no Bolts quasiregular
singularities are absent \cite{Anderson}.

For simplicity we shall concentrate mainly on the four-dimensional case.
However, as we shall see in the last section, our results can be easily generalized to higher dimensional situations.

Consider the spherical Taub-NUT-dS solution, which is constructed as a
circle fibration over the sphere in de Sitter background: 
\begin{equation}
ds^{2}=V(\tau )(dt+2n\cos \theta d\phi )^{2}-\frac{d\tau ^{2}}{V(\tau )}%
+(\tau ^{2}+n^{2})(d\theta ^{2}+\sin ^{2}\theta d\phi ^{2})
\end{equation}%
where 
\begin{equation}
V(\tau )=\frac{\tau ^{4}+(6n^{2}-l^{2})\tau ^{2}+2ml^{2}\tau
-n^{2}(3n^{2}-l^{2})}{(\tau ^{2}+n^{2})l^{2}}  \label{VdSTN}
\end{equation}%
As noted in the previous section, there are two different approaches to
describe the thermodynamics of such solutions, namely the $\mathbb{C}$- and
the $\mathbb{R}$-approach depending on the various analytic continuations
that can be done.

\subsection{The $\mathbb{C}$-approach results}

In the $\mathbb{C}$-approach one analytically continues the coordinates in the $(t,\tau )$
sector such that the signature in this section becomes $(++)$ or $(--)$. One way
to accomplish this is to analytically continue the coordinate $t\rightarrow
iT$ and the nut charge parameter $n\rightarrow iN$. One obtains the metric: 
\begin{equation}
ds^{2}=-F(\tau )(dT-2N\cos \theta d\phi )^{2}-\frac{d\tau ^{2}}{F(\tau )}%
+(\tau ^{2}-N^{2})(d\theta ^{2}+\sin ^{2}\theta d\phi ^{2})
\end{equation}%
where now 
\begin{equation}
F(\tau )=\frac{\tau ^{4}-(6N^{2}+l^{2})\tau ^{2}+2ml^{2}\tau
-N^{2}(3N^{2}+l^{2})}{(\tau ^{2}-N^{2})l^{2}}
\end{equation}

The conserved mass associated with this solution is \cite{rick2,rick1} 
\begin{equation}
\mathcal{M}_{c}=-m \label{MconsC}
\end{equation}%
independently of whether the function $F(\tau )$ has any roots. Note that if
we take $m<0$, this will violate the maximal mass conjecture. When $F(\tau )$
has roots the parameters $\left( m,N\right) $ are constrained relative to
one another by additional periodicity requirements. Even in this case the
maximal mass conjecture can be violated \cite{rick2,rick1}.

Notice that indeed the signature of the metric becomes in this case $(--++)$. When analysing the singularity structure of such spaces we have to take into account the presence of Misner string singularities as well as the possible conical singularities in the $(T,\tau )$ sector. To eliminate the Misner string singularity we impose the condition that the coordinate $T$ has in general the periodicity $\frac{8\pi N}{q}$, where $q$ is a positive integer which will also determine the topological structure of these solutions (see also \cite{Jensen}). To see this, notice that regularity of the $1$-form $(dT-2N\cos\theta d\phi)$ is achieved once we set the periodicity of $t$ to be given compatible with the integrals of $2N\sin\theta d\theta\wedge d\phi$ over all $2$-cycles in the base manifold. In $4$-dimensions the base is $S^2$ thence the value of the integral is $8\pi N$. While one could simply consider the Bolt solution corresponding to $q=1$, if $q>1$ then the topology of the Bolt solution is in general that of an $R^2/Z_q$ fibration over $S^2$. In order to get rid of the conical singularities in the $(T,\tau )$ sector we require the coordinate $T$ be periodically identified with periodicity given by $\beta _{c}=\frac{4\pi }{|F^{\prime }(\tau _{c})|}$, where $\tau _{c}$ is a root of $F(\tau )$, i.e. $F(\tau _{c})=0$, provided that such a root exists. For consistency, we have to match the values of the two obtained periodicities and this yields the condition: 
\begin{equation}
\beta _{c}=\frac{4\pi }{|F^{\prime }(\tau _{c})|}=\frac{8\pi |N|}{q}
\label{R-period}
\end{equation}

After some algebra, it can be readily checked that in this case we obtain 
\begin{equation*}
m=m_{c}=-\frac{\tau _{c}^{4}-(l^{2}+6N^{2})\tau _{c}^{2}-N^{2}(l^{2}+3N^{2})%
}{2l^{2}\tau _{c}}
\end{equation*}%
where 
\begin{equation*}
\tau _{c}^{\pm }=\frac{ql^{2}\pm \sqrt{q^{2}l^{4}+48N^{2}l^{2}+144N^{4}}}{%
12|N|}
\end{equation*}%
In order to satisfy the condition $|\tau _{c}|>\left| N\right| $ we must
consider the positive sign in the preceding expression for $N>0$, and the
negative sign for $N<0$ and in the last case we should also require $q=1$.

Working at future infinity it can be shown that if function $F(\tau )$ has roots then we obtain the action, respectively the entropy: 
\begin{equation}
I_{c}=-\frac{\beta _{c}(\tau _{c}^{3}-3N^{2}\tau _{c}+m_{c}l^{2})}{2l^{2}},~~~~~~~S_{c}=\frac{\beta _{c}(\tau _{c}^{3}-3N^{2}\tau _{c}-m_{c}l^{2})}{2l^{2}}
\end{equation}
A more detailed account of the thermodynamics of these solutions can be
found in \cite{rick2,rick1}. An interesting path-integral study of these solutions aimed at their cosmological interpretations appeared in \cite{Jensen}.\footnote{We thank the anonymous referee for bringing out this paper to our attention.}  However, while $\tau = \tau _{c}^{+}$ is the
largest root of the upper branch solutions, it can easily be checked that for all
lower branch solutions the function $F(\tau ,\tau _{c}^{-})$ has always two
roots $\tau _{1}$ and $\tau_{2}$ such that $\tau _{1}<\tau _{c}^{-}<\tau
_{2} $. Hence these solutions, though they \textit{apparently} respect the
first law of thermodynamics, are not valid dS-bolt solutions. Rather they
are the analytic continuation of lower-branch AdS-bolt solutions, as we shall see below. Furthermore, they have no counterpart in the $\mathbb{R}$-approach, as we shall also see.

\subsection{The $\mathbb{R}$-approach results}

In the $\mathbb{R}$-approach one does not analytically continue either the
coordinates or the parameters in the metric. Instead one directly uses the
metric in the Lorentzian signature 
\begin{equation}
V(\tau )=\frac{\tau ^{4}+(6n^{2}-l^{2})\tau ^{2}+2ml^{2}\tau
-n^{2}(3n^{2}-l^{2})}{(\tau ^{2}+n^{2})l^{2}}
\end{equation}%
where $n$ is the nut charge. In \cite{rick2} the conserved mass was found
to be 
\begin{equation}
\mathcal{M}_{r}=-m
\end{equation}%
for arbitrary values of the parameters $\left( m,n\right) $. Again, setting $%
m<0$ will violate the maximal mass conjecture.

Notice that in this case the coordinate $t$ parameterizes a circle fibered
over the $2$-sphere with coordinates $(\theta ,\phi )$. In the $\mathbb{R}$
-approach one imposes directly the periodicity condition on the spacelike
coordinate $t$: 
\begin{equation}
\beta _{r}=\frac{4\pi }{|V^{\prime }(\tau _{r})|}=\frac{8\pi n}{k}
\label{betaro}
\end{equation}%
for points where $V(\tau _{r})=0$ (provided that $\tau _{r}$ exists) with $k$ a
positive integer. Since these surfaces are two-dimensional (they are the
usual Lorentzian horizons) we shall still refer to as `bolts'. From the
above condition one obtains: 
\begin{equation}
m=m_{r}=-\frac{\tau _{r}^{4}+(6n^{2}-l^{2})\tau _{r}^{2}+n^{2}(l^{2}-3n^{2})%
}{2l^{2}\tau _{r}}
\end{equation}%
where now 
\begin{equation}
\tau _{r}^{\pm }=\frac{kl^{2}\pm \sqrt{k^{2}l^{4}-144n^{4}+48n^{2}l^{2}}}{12n}\label{realboltR}
\end{equation}%
In order to have real roots we must impose the condition that the
discriminant above be positive. This restricts the possible values of $n$
and $l$ such that: 
\begin{equation*}
|n|\leq \left( \frac{2+\sqrt{4+k^{2}}}{12}\right) ^{\frac{1}{2}}l
\end{equation*}

Provided that such $\tau _{r}$ exists we obtain 
\begin{equation}
I_{r}=-\frac{\beta _{r}(m_{r}l^{2}+\tau _{r}^{3}+3n^{2}\tau _{r})}{2l^{2}},~~~~~~~S_{r}=-\frac{\beta _{r}(m_{r}l^{2}-3n^{2}\tau _{r}-\tau _{r}^{3})}{2l^{2}}
\end{equation}%
for the action and entropy respectively. Although the
additional periodicity constraint (\ref{R-period}) imposes further
restrictions, the maximal mass conjecture can again be violated for certain
values of the parameters \cite{rick2,rick1}. It can be readily checked the
first law of thermodynamics is satisfied for both the upper ($\tau _{r}^{+}$%
)and lower ($\tau _{r}^{-}$) branch solutions. However, as in the previous
situation using the $\mathbb{C}$-approach, the function $V(\tau )$ will always have
two roots $\tau _{1}$ and $\tau _{2}$ such that $\tau _{1}<\tau
_{r}^{-}<\tau _{2}$. That the first law holds for the lower branch is a
direct consequence of the fact that this solution can be regarded as the
analytic continuation $l\rightarrow il$ of one of the Bolt solutions in the
Taub-NUT AdS case (see equation (\ref{rboltsads}) bellow), for which the
first law holds.

In what follows we shall restrict our analysis only to the upper branch
solutions given by $\tau _{r}^{+}$.

\subsection{From the $\mathbb{C}$-approach to the $\mathbb{R}$-approach}

As we have seen above we have two apparently distinct\ approaches for
describing thermodynamics of Taub-NUT-dS spaces. In the $\mathbb{C}$%
-approach we consider the analytic continuation of the spacelike $t$
coordinate and of the nut parameter. While this procedure will generally
lead to a `wrong signature' metric, this is simply a consequence of the fact
that we work in the region outside the cosmological horizon; the metric
inside the cosmological horizon has Euclidian signature. The signature in
the $(T,\tau )$ sector is in this case $(-,-)$ and we impose a periodicity of the
coordinate $T$ in order to get rid of the possible conical singularities in this sector. When
matched with the periodicity required by the absence of the Misner string singularity, this will fix, in general, the form of the mass parameter and the location of the nuts and bolts. We can now use the counterterm method to compute conserved quantities and study the thermodynamics of these solutions.

On the other hand, in the $\mathbb{R}$-approach we work directly with the
fields defined in the Lorentzian signature section. There is a periodicity
of the spacelike coordinate $t$ that appears from the requirement that there
are no Misner string singularities; however since the signature in the $(t,\tau )$ sector is now $(+,-)$ there are no conical singularities to be eliminated, so that there is no apparent reason to impose an extra condition as in (\ref{betaro}). Indeed, in the absence of the Hopf-type fibration the
coordinate $t$ is not periodic.

However, there is no a-priori obstruction in formally satisfying eq. (\ref{betaro}), and then using\footnote{
Notice that we use the metric in the Lorentzian signature} the counterterm method to compute the conserved quantities and study the thermodynamics of these solutions as it was done in refs. \cite{rick2,rick1}. We shall show in what follows that in general the $\mathbb{R}$-approach results are just the analytic continuation of the $\mathbb{C}$-approach results and vice-versa. We shall later show that this affords a physical interpretation of the thermodynamics of Lorentzian nut-charged spacetimes.

To motivate this claim we consider the followings. In order to obtain a
Euclidian signature (positive or negative definite) in the $(t,\tau )$
sector we must perform the analytic continuations $t\rightarrow iT$ and $%
n\rightarrow iN$. However, since the function $V(\tau )$ depends only on $n$
and not on $t$ its analytic continuation will be given by: 
\begin{equation}
F(\tau )=\frac{\tau ^{4}-(6N^{2}+l^{2})\tau ^{2}+2ml^{2}\tau
-N^{2}(3N^{2}+l^{2})}{(\tau ^{2}-N^{2})l^{2}}  \label{Ftau}
\end{equation}%
It is readily seen that using these analytical continuations we obtain the metric used
in the $\mathbb{C}$-approach. The key point to notice here is that we can go
back to the Lorentzian section by again employing the analytic continuations 
$T\rightarrow it$ and $N\rightarrow in$. Since only even powers of $n$
appear we are guaranteed that the above continuations will take us from the
Lorentzian section to the `Euclidian' one and back.

In the $\mathbb{C}$-approach it makes sense to consider the removal of
conical singularities in the $(T,\tau )$ sector (since the signature of the
metric in that sector is $(--)$), as well as to match this periodicity
condition with the one arising by requiring the absence of Misner string
singularities. Hence we are fully entitled to impose the condition: 
\begin{equation}
\beta _{c}=\frac{4\pi }{|F^{\prime }(\tau _{c})|}=\frac{8\pi N}{q}
\label{cper}
\end{equation}%
where $q$ is a positive integer and $\tau _{c}$ is such that $F(\tau _{c})=0$.

Let us consider now the effect of the analytic continuations $T\rightarrow
it $ and $N\rightarrow in$ on the above condition. Since nothing depends on $%
T$ explicitly, all that matters is the effect of the analytic continuation of the nut
charge. If we continue $N\rightarrow in$ we obtain $V(\tau )=F(\tau
)|_{N=in} $. Thus in the above periodicity condition we obtain 
\begin{equation}
\left( \frac{4\pi }{|F^{\prime }(\tau _{c})|}\right) _{N=in}=\frac{4\pi }{%
|V^{\prime }(\tau _{r})|}
\end{equation}
where now $\tau _{r}$ is such that $V(\tau _{r})=F(\tau _{c})=0$. Hence $%
\beta _{r}=(\beta _{c})_{N=in}$. However, as we can see from the second
equality in (\ref{cper}) we can consistently analytically continue $%
N\rightarrow in$ only if we also continue $q\rightarrow ik$. Again this
assures us that $\beta _{r}=(\beta _{c})_{N=in}$ and that it is real.

Thus the prescription to get the $\mathbb{R}$-results from the $\mathbb{C}$%
-results is as follows: using the $\mathbb{C}$-results perform the analytic
continuations $T\rightarrow it$, $N\rightarrow in$ \textit{and} $%
q\rightarrow ik$. A naive analytic continuation only of $T$ and $N$ but
without continuing $q$ is simply inconsistent\footnote{%
If we rewrite eq. (\ref{cper}) by taking the absolute value of both sides
then the continuation $q\rightarrow ik$ while no longer necessary, is still
permitted.}: from eq. (\ref{cper}) the left hand side remains real while the
right hand side becomes complex!

Let us check this prescription by obtaining the $\mathbb{R}$-results
starting from the $\mathbb{C}$-results for the thermodynamic quantities
given above. Consider first the location of the bolts in the $\mathbb{C}$%
-approach: 
\begin{equation}
\tau _{c}^{+}=\frac{ql^{2}+\sqrt{q^{2}l^{4}+48N^{2}l^{2}+144N^{4}}}{12N}
\end{equation}%
If we perform $N\rightarrow in$ and $q\rightarrow ik$ we obtain $\tau
_{c}\rightarrow \tau _{r}$ where: 
\begin{equation}
\tau _{r}^{+}=\frac{kl^{2}+\sqrt{k^{2}l^{4}-144n^{4}+48n^{2}l^{2}}}{12n}
\end{equation}%
This is indeed the location of the `bolt' in the $\mathbb{R}$-approach as we can see from (\ref{realboltR}).

Performing the analytic continuations $N\rightarrow in$ and $%
\tau^{+}_{c}\rightarrow \tau^{+}_{r}$ in the expression for the mass
parameter in the $\mathbb{C}$-approach we obtain $m_{c}\rightarrow m_{r}$
where: 
\begin{equation}
m_{r}=-\frac{(\tau^{+}_{r})^{4}+(6n^{2}-l^{2})(%
\tau^{+}_{r})^{2}+n^{2}(l^{2}-3n^{2})}{2l^{2}\tau^{+}_{r}}
\end{equation}
which again is the mass parameter from the $\mathbb{R}$-approach. Notice
that if we naively analytically continue $N\rightarrow in$ and ignore the
condition $q\rightarrow ik$ we obtain imaginary values for the corresponding
results in the $\mathbb{R}$-approach. However both the above analytic
continuations conspire to always produce real quantities in the final
results.

A closer look at the expressions for the action, conserved mass and the
entropy in the $\mathbb{C}$-approach shows that if we perform the
continuations $N\rightarrow in$, $\tau^{+}_{c}\rightarrow \tau^{+}_{r}$ and $%
m_{c}\rightarrow m_{r}$ we obtain the respective expressions from the $%
\mathbb{R}$-approach.

\section{No dS NUTs}

It is known that in the asymptotically AdS/flat case, besides the usual Taub-Bolt
solutions, one can also obtain the so-called Taub-Nut solutions. For these
solutions the fixed-point set of the Killing vector $\frac{\partial }{
\partial T}$ is zero-dimensional.

Superficially, a similar situation appears to hold in the $\mathbb{C}$-approach in dS backgrounds \cite%
{rick2}. This can happen only if $\tau _{c}=N$ in the above equations, that
is if $F(\tau _{c}=N)=0$ and also 
\begin{equation}
\frac{4\pi }{|F^{\prime }(\tau _{c}=N)|}=\frac{8\pi N}{q}  \label{Nutperiod}
\end{equation}%
are satisfied. Although such an equation has solutions, we find that the
situation is more somewhat complicated than previously described in ref. \cite{rick2}.

Solving (\ref{Nutperiod}) we find $q=1$, \textit{i.e.} the periodicity of the $T$ coordinate is $8\pi N$, while the mass parameter becomes: 
\begin{equation}
m_{c}=\frac{N(l^{2}+4N^{2})}{l^{2}}  \label{mdsNUT}
\end{equation}
and indeed $\tau _{c}=N$ is a fixed point set of zero dimensionality. However it is not the largest root of the function $F(\tau )$ as given in (\ref{Ftau}). Instead this nut is contained within a
larger cosmological `bolt' horizon located at $\tau =\tau _{ch}=\sqrt{4N^{2}+l^{2}}-N$.
In this sense there are no dS NUTs, \textit{i.e.} no outmost cosmological horizons that are dimension zero fixed point sets of the Killing vector $\frac{\partial }{\partial T}$.

Note, however that if we insert this value for the mass parameter into eqs. (\ref{MconsC}) and subsequently (\ref{GDuhem}) we obtain the action and respectively the entropy:
\begin{equation}
I_{c}=-\frac{4\pi N^{2}(l^{2}+2N^{2})}{l^{2}},~~~~~~~S_{c}=-\frac{4\pi N^{2}(l^{2}+6N^{2})}{l^{2}}
\end{equation}
These values correspond to those derived for the Taub-Nut-dS solution studied in \cite{rick2}, and it is straightforward to show that the first law of thermodynamics is obeyed.

However the physical interpretation of this solution is not as a Taub-Nut in dS background,
since the use of such formulae is predicated on $\tau _{c}=N$ being the largest root of $F$. Rather this solution is the AdS-NUT under the analytic continuation $l\rightarrow il$ (see (\ref{cnutentropyads}) below). We shall discuss the corresponding solution when we address the AdS case.

It is straightforward to show that the putative `bolt' solution, with $\tau_{ch}=\sqrt{4N^{2}+l^{2}}-N$, yields an entropy that does not respect the first law of thermodynamics. This presumably is a consequence of the fact that we eliminated the conical singularity at the root $\tau _{c}=N$, by fixing the periodicity of the Euclidian time $T$ to be $8\pi N$, while leaving a conical singularity that can not be eliminated at the outer root $\tau _{ch}$! However, upon further inspection we find that if we choose to eliminate the
conical singularity at the outer root and fix the periodicity of the Euclidian time coordinate to be $\beta _{c}=\frac{4\pi }{|F^{\prime }(\tau_{ch})|}$, we still obtain a singular space. This is because the Misner
string singularity cannot be simultaneously eliminated unless we impose a relationship between the nut charge and the cosmological constant.\footnote{In this case we might be able to recover a regular Euclidian instanton, having the $CP^2$ topology: with a nut at $\tau_c=N$ and a bolt at $\tau_{ch}$.} Furthermore, the entropy (as computed via the counterterm method) does not satisfy the first law. The difficulties in ascribing a consistent thermodynamic interpretation to this solution make its physical relevance a dubious prospect.

\section{Taub-NUT solutions in $AdS/$flat backgrounds}

Motivated by the results of the previous sections we shall now extend
our prescription to describe the thermodynamics of the Taub-NUT solutions in 
$AdS$ or flat backgrounds. To our knowledge, the thermodynamics of such
solutions has been discussed only in the $\mathbb{C}$-approach (\textit{i.e.}
in Euclidian regime) in \cite{mann1,Hawking2,myers1,rick3,myers2}. To begin with, let us
recall the metric of the Taub-NUT AdS solution in four dimensions: 
\begin{equation}
ds^{2}=-V(r)(dt-2n\cos \theta d\phi
)^{2}+V^{-1}(r)dr^{2}+(r^{2}+n^{2})d\Omega ^{2}  \label{ads1}
\end{equation}%
where $d\Omega ^{2}=d\theta ^{2}+\sin ^{2}\theta d\phi ^{2}$ is the metric
on the sphere $S^{2}$ and 
\begin{equation}
V(r)=\frac{r^{4}+(l^{2}+6n^{2})r^{2}-2mrl^{2}-n^{2}(l^{2}+3n^{2})}{%
l^{2}(n^{2}+r^{2})}  \label{ads1V}
\end{equation}%
This metric is a solution of the vacuum Einstein field equations with
negative cosmological constant $\lambda =-\frac{3}{l^{2}}$. In the limit $%
l\rightarrow \infty $ it reduces to the usual asymptotically (locally) flat
Taub-NUT solution.

In the $\mathbb{C}$-approach we analytically continue the time coordinate $%
t\rightarrow iT$ and the nut charge $n\rightarrow iN$. We obtain a Euclidian
signature metric of the form: 
\begin{equation}
ds^{2}=F(r)(dT-2N\cos \theta d\phi
)^{2}+F^{-1}(r)dr^{2}+(r^{2}-N^{2})d\Omega ^{2}  \label{adsC}
\end{equation}%
where 
\begin{equation}
F(r)=\frac{r^{4}+(l^{2}-6N^{2})r^{2}-2mrl^{2}+N^{2}(l^{2}-3N^{2})}{%
l^{2}(r^{2}-N^{2})}  \label{adsC2}
\end{equation}%
When discussing the singularity structure of these spaces we must impose two regularity conditions. First, removal of the Misner string singularities leads us to periodically identify the coordinate $T$ with period $\frac{8\pi N}{q}$, where $q$ is a non-negative integer. Now, if we match this value with the periodicity obtained by removing the conical singularities at the roots $r_c$ of the function $F(r)$ we obtain in general 
\begin{equation}
\beta _{c}=\frac{4\pi }{F^{\prime }(r_{c})}=\frac{8\pi N}{q}
\end{equation}%
 Again we have two distinct cases to consider: in the Taub-Nut solution we impose $r_{c}=N$ (which
makes the fixed-point set of the isometry $\partial _{T}$ zero-dimensional), whereas for the bolt solutions $r_{c}=r_{b\pm }>N$ (for which the fixed-point set of $\partial _{T}$ is two-dimensional).

For the Taub-Nut solution $r_{c}=N$, the periodicity of the coordinate $T$ is found to be $8\pi N$, \textit{i.e.} $q=1$, and the value of the mass parameter is: 
\begin{equation*}
m_{c}=\frac{N(l^{2}-4N^{2})}{l^{2}}
\end{equation*}%
The action and entropy are: 
\begin{equation}
I_{c}=\frac{4\pi N^{2}(l^{2}-2N^{2})}{l^{2}},~~~~~~~S_{c}=\frac{4\pi N^{2}(l^{2}-6N^{2})}{l^{2}}  \label{cnutentropyads}
\end{equation}%
while the specific heat $C=-\beta _{c}\partial _{\beta _{c}}S$ is given by: 
\begin{equation*}
C_{c}=\frac{8\pi N^{2}(12N^{2}-l^{2})}{l^{2}}
\end{equation*}

Notice that the energy $\frak{M}=m$ becomes negative if $N>\frac{l}{2}$ while the action
becomes negative for $N>\frac{l}{\sqrt{2}}$ . When this latter inequality is
saturated we recover the Euclidian AdS spacetime. The entropy is negative if $N>%
\frac{l}{\sqrt{6}}$ while for $N<\frac{l}{\sqrt{12}}$ the specific heat
becomes negative, which signals thermodynamic instabilities. Therefore, it
has been argued in ref. \cite{myers2} that in order to obtain physically relevant
solutions with both positive entropy and specific heat one should restrict
the values of the nut charge such that: 
\begin{equation}
\frac{l}{\sqrt{12}}\leq N\leq \frac{l}{\sqrt{6}}
\end{equation}

The other possibility corresponds to the Bolt solutions, for which $r=r_{c}>N $. In this case the periodicity of the coordinate $T$ is $\frac{8\pi N}{q}$, with $q$ a non-negative integer. While value $q=1$ is somehow singled out as it leads to identical periodicity with the one from the Nut solution, other values $q>1$ are allowed as well. For a general $q$ the topology on the boundary is that of a lens space $S^3/Z_q$, while the topology of the Bolt is in general that of an $R^2/Z_q$-fibration over $S^2$. The value of the mass parameter is 
\begin{equation}
m_{c}=\frac{r_{c}^{4}+(l^{2}-6N^{2})r_{c}^{2}+N^{2}(l^{2}-3N^{2})}{%
2l^{2}r_{c}}  \label{massc}
\end{equation}
while the location of the bolts is given by: 
\begin{equation}
r_{c}=\frac{ql^{2}\pm \sqrt{q^{2}l^{4}-48N^{2}l^{2}+144N^{4}}}{12N}
\label{rcads}
\end{equation}

Notice that the condition $r_{c}>N$ restricts the values of the NUT charge
parameter $N$ such that (for $q=1$): 
\begin{eqnarray}
N\leq \left( \frac{1}{6}-\frac{\sqrt{3}}{12}\right) ^{\frac{1}{2}}l
\end{eqnarray}
For the bolt solutions the action is given by \cite{rick3} 
\begin{eqnarray}
I_{c}=-\frac{\pi (r_{c}^{4}-l^{2}r_{c}^{2}+N^{2}(3N^{2}-l^{2}))}{%
3r_{c}^{2}-3N^{2}+l^{2}}
\end{eqnarray}
and the entropy is 
\begin{eqnarray}
S_{c}=\frac{\pi (3r_{c}^{4}+(l^{2}-12N^{2})r_{c}^{2}+N^{2}(l^{2}-3N^{2}))}{%
3r_{c}^{2}+l^{2}-3N^{2}}
\end{eqnarray}

Note that the properties of the bolt solution with $r>r_{b+}$ are very
different from those of the bolt solution with $r>r_{b-}$. It can be shown
that the upper branch solution $r>r_{b+}$ is thermally stable whereas the
lower branch $r>r_{b-}$ is thermally unstable \cite{rick3,myers2}.

We shall now apply our prescription to convert all $\mathbb{C}$-results to
the corresponding results in the $\mathbb{R}$-approach, for which the metric
used has the Lorentzian signature.

Since we do not perform any analytic continuations in the $\mathbb{R}$%
-approach, the metric that we use is given by (\ref{ads1}). The periodicity
condition for the coordinate $t$ is then given by: 
\begin{equation}
\frac{4\pi }{|V^{\prime }(r_{r})|}=\frac{8\pi n}{k}
\end{equation}%
where $V(r_{r})=0$ and $k$ is a positive integer.

\section{Euclidian to Lorentzian}

Before we plunge into the details of the Euclidian to Lorentzian transition
by analytical continuation it is necessary first to discuss what is to
become the first law of the thermodynamics in this process.

\subsection{When is the first law of thermodynamics satisfied?}

\label{first law}

It has been recently argued in \cite{Micky1a} that there is a breakdown of the entropy/area relationship for nut-charged AdS-spacetimes and that this result does not depend on the removal of Misner string singularities (if present) but rather is entirely a consequence of the first law of thermodynamics.

In the Euclidian sector (or the $\mathbb{C}$-approach) the argument goes as
follows: using the counterterm method for a general bolt located at $r=r_{c}$
we compute 
\begin{equation*}
I_{c}=\frac{\beta _{c}}{2l^{2}}(l^{2}m_{c}+3N^{2}r_{c}-r_{c}^{3})
\end{equation*}
for the action, where $\beta _{c}=\frac{4\pi}{|F^{\prime}(r_c)|}$ is the
periodicity of the Euclidian time coordinate and $r_{c}$ is the biggest root
of $F(r)$ given in (\ref{adsC}). This fixes the value of the mass parameter to be
that given by (\ref{massc}). Using the boundary stress-energy tensor we can
compute the conserved mass for this solution as being given by $\mathfrak{M}=m_{c}$ \cite{myers2,Micky1a}. We define now the entropy $S_{c}=\beta _{c}m_{c}-I_{c}$ by using the Gibbs-Duhem relation. It is easy to see that in order for the first law of thermodynamics $dS_{c}=\beta
_{c}dm_{c}$ to hold in this case we must have: 
\begin{eqnarray}
m_{c}=\partial _{\beta _{c}}I_{c}
\end{eqnarray}
For generic values of $r_{c}$ we find that the above relation is not satisfied in general. However, if we assume a functional dependence $r_{c}=r_{c}(N)$ then the first law is satisfied if and only if $r_{c}$ is
given by (\ref{rcads}) where now $q$ is a constant of integration or $r_c=\pm n$.

We can see now that we are guaranteed to have satisfied the first law of
thermodynamics for the Nut and Bolt solutions in
AdS backgrounds, even though no Misner-string singularities have been
explicitly removed. We also find that using the expressions from (\ref{rcads}) we
obtain $\beta=\frac{8\pi|N|}{q}$. Now, removal of Misner-string singularities
forces the parameter $q$ to be an integer, but this is not required in order
to satisfy the first law.

Let us consider next the restrictions imposed by the first law of
thermodynamics in the $\mathbb{R}$-approach, \textit{i.e.} in the Lorentzian
solutions. Using the counterterm method for the Lorentzian solution given
in (\ref{ads1}) with a 'bolt' located at $r_{r}$, which is a root of (\ref{ads1V}), we obtain the action: 
\begin{equation*}
I_{r}=\frac{\beta _{r}}{2l^{2}}(l^{2}m_{r}-3n^{2}r_{r}-r_{r}^{3})
\end{equation*}%
Here we set $\beta _{r}=\frac{4\pi}{|V^{\prime}(r_r)|}$ to be the
periodicity of the time coordinate, though there is no direct justification
for this\footnote{%
Note however that the removal of the Misner string singularity in the
Lorentzian metric forces the time coordinate to be periodic.}. We find the
value of the mass parameter $m_{r}$ to be 
\begin{eqnarray}
m_{r}=\frac{r_{r}^{4}+(l^{2}+6n^{2})r_{r}^{2}-n^{2}(l^{2}+3n^{2})}{%
2l^{2}r_{r}}  \label{mrboltsads}
\end{eqnarray}
Using the boundary stress-energy tensor we can compute the conserved mass
for this solution as being given by $\mathfrak{M}=m_{r}$. We can now define
the entropy $S_{r}=\beta _{r}m_{r}-I_{r}$ by using the Gibbs-Duhem relation.
It is easy to see that in order for the first law of thermodynamics $dS_{r}=\beta _{r}dm_{r}$ to hold in this case we must have: 
\begin{equation*}
m_{r}=\partial _{\beta _{r}}I_{r}
\end{equation*}
Again, for generic values of $r_{r}$ we find that the above relation is not
satisfied in general. However if we assume a functional dependence $%
r_{r}=r_{r}(n)$ then the first law is satisfied if and only if: 
\begin{equation}
r_{r}=\frac{kl^{2}\pm \sqrt{k^{2}l^{4}-48n^{2}l^{2}-144n^{4}}}{12n}
\label{rr2}
\end{equation}%
where $k$ is a constant of integration. As we shall see in the next section,
this is precisely the location of the Lorentzian bolt solutions, when $k$ is an integer. The
first law of thermodynamics will be automatically satisfied for these solutions. It is
interesting to note that using the expressions from (\ref{mrboltsads}) and (\ref{rr2}) we obtain $\beta=\frac{8\pi|n|}{k}=\frac{4\pi}{|V^{\prime}(r_r)|}$. If we impose the further requirement that Misner string singularities be removed then $k$ must be an integer.

\subsection{The Bolt case}

Let us consider now the analytic continuation of the bolt solutions from the 
$\mathbb{C}$-approach. In this case we perform the analytic continuations $T\rightarrow it$, $N\rightarrow in$ together with $q\rightarrow ik$. From (\ref{rcads}) we obtain the location of the Lorentzian `bolts' at: 
\begin{equation}
r_{r}=\frac{kl^{2}\pm \sqrt{k^{2}l^{4}-48n^{2}l^{2}-144n^{4}}}{12n}
\label{rboltsads}
\end{equation}
the value of the mass parameter is: 
\begin{equation}
m_{r}=\frac{r_{r}^{4}+(l^{2}+6n^{2})r_{r}^{2}-n^{2}(l^{2}+3n^{2})}{%
2l^{2}r_{r}}
\end{equation}
while the periodicity of the time coordinate $t$ is given by $\beta _{r}=\frac{8\pi n}{k}$.

In order to obtain real values for $r_{r}$ we must require that the
discriminant is non-negative. This leads to the condition: 
\begin{equation}
n\leq n_{max}=\left( \frac{\sqrt{4+k^{2}}-2}{12}\right)^{\frac{1}{2}}l
\end{equation}
Then there is a maximum value $n_{max}$ of the nut charge for which the bolt solutions are physically acceptable. This means that below a certain temperature $T_{min}=\frac{k}{8\pi n_{max}}$ the bolt solutions do not exist.

If we analytically continue the action and the entropy of the bolt solutions
we obtain: 
\begin{eqnarray}
I_{r}=-\frac{\pi (r_{r}^{4}-l^{2}r_{r}^{2}+n^{2}(3n^{2}+l^{2}))}{%
3r_{r}^{2}+3n^{2}+l^{2}}
\end{eqnarray}
respectively 
\begin{eqnarray}
S_{r}=\frac{\pi (3r_{r}^{4}+(l^{2}+12n^{2})r_{r}^{2}-n^{2}(l^{2}+3n^{2}))}{%
3r_{r}^{2}+l^{2}+3n^{2}}
\end{eqnarray}
The specific heats can be computed using $C=-\beta \partial
_{\beta}S=-n\partial _{n}S$ ; for brevity we shall not list here their
explicit expressions.

\begin{figure}[tbp]
\centering           
\begin{minipage}[c]{.45\textwidth}
         \centering
         \includegraphics[width=\textwidth,angle=0,keepaspectratio]{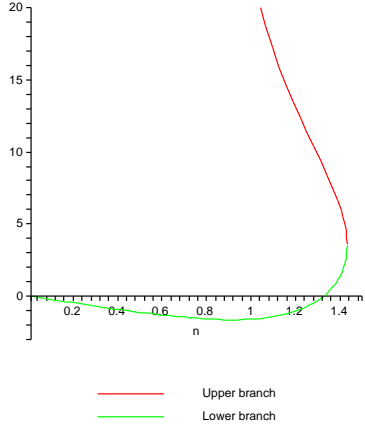}
\end{minipage}
\caption{Plot of the upper ($r _{b}=r _{b+}$) and lower ( $r _{b}=r_{b-}$)
TB masses (for $k=10$, $l=\protect\sqrt{3}$).}
\label{Boltmass}
\end{figure}

In figure \ref{Boltmass} we plot the masses of the upper branch ($r>r_{b+}$)
and the lower branch ($r>r_{b-}$) solutions as a function of the nut
parameter $n$. We can see that there is a range for the nut charge for which
the mass of the lower branch solution becomes negative, while the mass of
the upper branch solution is always positive.

\begin{figure}[tbp]
\centering           
\begin{minipage}[c]{.45\textwidth}
         \centering
         \includegraphics[width=\textwidth,angle=0,keepaspectratio]{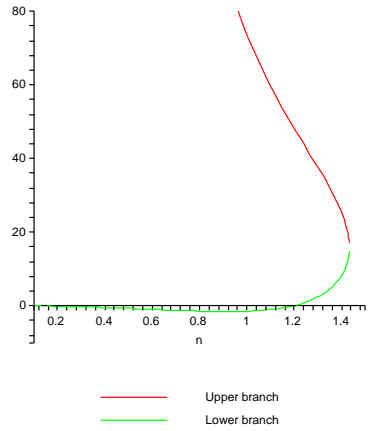}
         \end{minipage}
\caption{Plot of the upper ($r_{b}=r_{b+}$) and lower ( $r_{b}=r_{b-}$) TB
entropies (for $k=10$, $l=\protect\sqrt{3}$).}
\label{Boltentropy}
\end{figure}

We plot the entropy as a function of the nut charge in figure \ref{Boltentropy}, including the lower branch solutions. As is obvious from this figure, the entropy for the lower branch does become negative if $n<1.19355$. The entropy for the upper branch solutions is always positive.

\begin{figure}[tbp]
\centering           
\begin{minipage}[c]{.45\textwidth}
         \centering
         \includegraphics[width=\textwidth,angle=0,keepaspectratio]{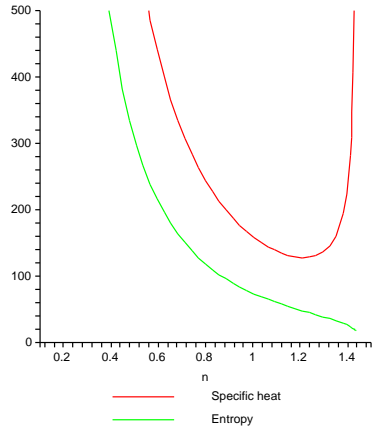}

\end{minipage}
\caption{Plot of the upper branch bolt entropy and specific heat (for $k=10$, $l=\protect\sqrt{3}$). }
\label{UpperSC}
\end{figure}

\begin{figure}[tbp]
\centering           
\begin{minipage}[c]{.45\textwidth}
         \centering
         \includegraphics[width=\textwidth,angle=0,keepaspectratio]{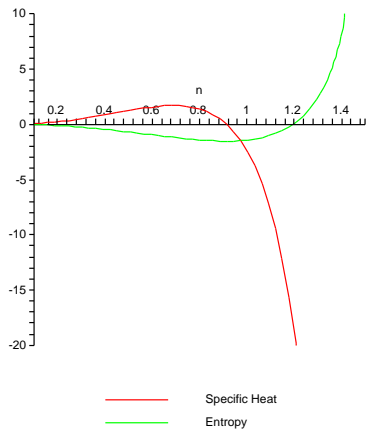}
         \end{minipage}
\caption{Plot of the lower branch bolt entropy and specific heat (for $k=10$, $l=\protect\sqrt{3}$). }
\label{LowerSC}
\end{figure}

In figure \ref{UpperSC} we plot the entropy and the specific heat versus the
nut charge for the upper branch solutions. In this case the entropy and the
specific heat are always positive. In figure \ref{LowerSC} we plot the
entropy and the specific heat as a function of the nut charge for the lower
branch bolt solutions. We can see again that the entropy is negative if $%
n<1.19355$ while the specific heat is positive if $n<0.91338$ and negative
otherwise, implying that the Lorentzian version of the lower branch
solutions is thermally unstable. In both cases the specific heat diverges
near $T=T_{min}$ (or $n=n_{max}$).

\subsection{The Nut case}

As we have seen above in the $\mathbb{C}$-approach, besides the usual Bolt
solutions, one can also obtain the so-called Nut-solutions. For these
solutions the fixed-point set of the Killing vector $\frac{\partial }{%
\partial T}$ is zero-dimensional. This can happen only if $\tau _{c}=N$ in
the above equations, that is if $F(\tau _{c}=N)=0$ and also 
\begin{equation}
\frac{4\pi }{|F^{\prime }(\tau _{c}=N)|}=\frac{8\pi N}{q}
\end{equation}
are satisfied. Recall that $q=1$ for the Nut solution in the $\mathbb{C}$-approach.

However, since we are interested in the analytic continuations that could
take the $\mathbb{C}$-results to the $\mathbb{R}$-results we shall slightly
modify our ansatz using the lesson learned in dealing with the Bolt cases.
Namely, instead of focussing on $\tau _{c}=N$ (which clearly becomes
imaginary when we analytically continue $N\rightarrow in$) we shall look for
a solution of the form $\tau _{c}=pN$ where $p$ is a positive real number.\footnote{We thank the anonymous referee for useful remarks on this point.} Then the usual Taub-Nut solution in the $\mathbb{C}$-approach corresponds to $p=1$, while other values of $p>1$ correspond to Bolt-type solutions.

The limit $p=1$ must be treated with special care since in the Nut solution $\tau _{c}=N$ is a double root of the numerator of $F(r)$, while in the Bolt case we assume that $\tau_c$ is a single root. The difference arises when computing the periodicity $\beta_{c}=\frac{4\pi}{|F^{\prime }(pN)|}$; accounting for the double root, it turns out that one should multiply by $2$ the result from the Bolt case in order to recover the correct periodicity of the Nut.

It is easy to check now that the above conditions will fix the periodicity
of the coordinate $T$ to be $\beta_c=\frac{8\pi N}{q}$ where now $q$ is a
complicated function of $p$, $l$ and $N$ while the value of the mass
parameter is given by 
\begin{eqnarray}
m_{c}&=&\frac{N[(1+p^2)l^2+(p^4-6p^2-3)N^2]}{2pl^2}
\end{eqnarray}

Using the counterterm method, for a bolt located at $r_{c}=pN$ we obtain: 
\begin{eqnarray}
I_{c} &=&\frac{\pi N^{2}[(p^{2}+1)l^{2}-(p^{4}+3)N^{2}]}{%
l^{2}+3N^{2}(p^{2}-1)}  \notag \\
S_{c} &=&\frac{\pi N^{2}[(p^{2}+1)l^{2}-3(p^{4}+3)N^{2}]}{%
l^{2}-3N^{2}(p^{2}-1)}
\end{eqnarray}
Notice that in the limit $p\rightarrow 1$ one recovers the previous
expressions for the action and respectively the entropy of the Taub-Nut-AdS solution (\ref{cnutentropyads}) up to the factor of $2$, as explained above.

Let us apply now our prescription from going from the $\mathbb{C}$-approach
to the $\mathbb{R}$-approach. In this case we shall analytically continue $%
N\rightarrow in$ and also $q\rightarrow ik$ (which in the Nut case corresponds in fact to $p\rightarrow -ip$). Then the location of the nut becomes $\tau _{r}=pn$ in the $\mathbb{R}$-approach, yielding a real value for $\tau _{r}$, while the value of the mass parameter is also real 
\begin{equation}
m_{r}=\frac{n[(p^{4}+6p^{2}-3)n^{2}-l^{2}(1-p^{2})]}{2pl^{2}}
\end{equation}
Now the periodicity of the coordinate $t$ is given by $\frac{8\pi n}{|k|}$,
with $k=(q)_{p=-ip}$. For $p=1$ we obtain: 
\begin{equation}
k=2\left( \frac{6n^{2}}{l^{2}}+1\right)  \label{knutb}
\end{equation}
while value of the mass parameter is $m_{r}=\frac{2n^{3}}{l^{2}}$, the
action is 
\begin{equation}
I_{r}=-\frac{4\pi n^{4}}{l^{2}+6n^{2}}
\end{equation}
However, since the location of the `bolt' $r=n$ is not of the form (\ref{rr2}%
) unless we assume a relationship between $n$ and $l$ we conclude that the
first law of thermodynamics is not satisfied for the Lorentzian Taub-Nut-AdS
solution in the $\mathbb{R}$-approach.

\subsection{The flat-space limit}

Leaving the more detailed study of the thermodynamics of the above solutions
for future work, let us now briefly discuss the case in which the
cosmological constant vanishes. Notice that this condition corresponds to $%
l\rightarrow \infty $ and in this limit we recover the Taub-NUT solutions in
a flat background. Special care must be taken when discussing the analytic
continuation from the Euclidian sector to the Lorentzian one. Let us
consider first the Lorentzian Taub-Nut-AdS solution. In the limit $l\rightarrow
\infty $ we obtain the action $I_{r}=0$, the conserved mass is also zero in
this limit. These results are in agreement with the expectation that the
only way to have $r=n$ as a root of the Lorentzian function 
\begin{equation*}
F(r)=\frac{r^{2}-2mr-n^{2}}{r^{2}+n^{2}}
\end{equation*}
is to take $m=0$.

In the bolt case we obtain by analytic continuation $r_{r}=\frac{2n}{k}$ and
the mass parameter is 
\begin{equation*}
m_{r}=n\frac{4-k^{2}}{4k}
\end{equation*}%
while the action and the entropy are given by: 
\begin{eqnarray}
I_{r} =\pi n^{2}\frac{4-k^{2}}{k^{2}},~~~~~S_{r} =\pi n^{2}\frac{4-k^{2}}{k^{2}},~~~~~
C_{r} =2\pi n^{2}\frac{k^{2}-4}{k^{2}}
\end{eqnarray}%
As in the Euclidian sector we have $q=1$ (since if $q>1$ then $r_{b}$ is
less than $n$) this will fix $k=1$ in the above relations. Further, note that
although these expressions satisfy the first law of thermodynamics, the
entropy and specific heat for the Lorentzian bolt in the flat spacetime have
opposite signs, which means that the solution is thermally unstable. This is not unexpected if we recall that the Taub-NUT solutions in flat background correspond to the lower-branch Taub-NUT-AdS solutions, which are thermodynamically unstable. 

\section{Higher dimensional Taub-NUT-dS spaces}

The above results can be easily extended to higher dimensional Taub-NUT
spacetimes. For simplicity we shall focus only on the asymptotically de
Sitter spacetimes. 

First, let us notice the absence of higher dimensional Taub-Nut-dS solutions, which is a result analogous with that stating the absence of hyperbolic nuts in $AdS$-backgrounds \cite{myers1,Micky1}. Quite generally we can see this by observing the behaviour of the function $F(\tau)$ near the root $\tau_c=N$. Since $F(\tau)$ takes negative values for points $\tau>\tau_c$ we deduce that there always exists a larger root of $F(\tau)$ that will contain the nut. Therefore, in our discussion we shall refer only to the higher-dimensional Bolt solutions. An analysis as the one performed in Section (\ref{first law}) assures us that the first law is satisfied in both approaches.

An analysis of the thermodynamics of the higher-dimensional Taub-NUT-dS spaces has been presented in \cite{rickreview}. It has been shown there that the thermodynamic behaviour of both the $\mathbb{R}$-approach and $\mathbb{C}$-approach quantities are qualitatively the same in $4s$-dimensions, a
behaviour that is distinct from the common behaviour in $4s+2$ dimensions\footnote{Here $s=1,2,...$.}. This means that spaces of dimensionality $8,12,16,...$ have the same qualitative thermodynamic behaviour as the four-dimensional case, while spaces of dimensionality $10,14,18,...$ have the same behaviour as the six-dimensional case. We shall now illustrate equivalence of the two approaches in six-dimensions; the other $4s+2$ higher-dimensional cases are analogous.

The Taub-NUT-dS metric in six dimensions, constructed over an $S^{2}\times S^{2}$ base is given by: 
\begin{equation*}
ds^{2}=V(\tau )(dt+2n\cos \theta _{1}d\varphi _{1}+2n\cos \theta
_{2}d\varphi _{2})^{2}-\frac{d\tau ^{2}}{V(\tau )}+(\tau ^{2}+n^{2})(d\Omega
_{1}^{2}+d\Omega _{2}^{2})
\end{equation*}%
where 
\begin{eqnarray}
V(\tau ) &=&\frac{3\tau ^{6}-(l^{2}-15n^{2})\tau
^{4}-3n^{2}(2l^{2}-15n^{2})\tau ^{2}+3n^{4}(l^{2}-5n^{2})+6ml^{2}\tau }{%
3(\tau ^{2}+n^{2})^{2}l^{2}}  \notag \\
d\Omega _{i}^{2} &=&d\theta _{i}^{2}+\sin ^{2}\theta _{i}d\varphi _{i}^{2}
\end{eqnarray}%
Removal of the Misner string singularities in the metric forces us to take $12\pi |n|$ as the periodicity of the time coordinate. Similar with the $4$-dimensional case, we shall impose an extra periodicity $\frac{4\pi}{|V^{\prime }(\tau _{r})|}$. Matching the two values leads to the condition: 
\begin{equation*}
\beta _{r}=\frac{4\pi }{|V^{\prime }(\tau _{r})|}=\frac{12\pi |n|}{k}
\end{equation*}
where $V(\tau _{r})=0$ and $k$ is a positive integer.\footnote{The parameter $k$ will determine the topology on the boundary $\tau\ra\infty$. For $k=1$ the boundary is $Q(1,1)$, which is the $5$-dimensional circle fibration over $S^2\times S^2$, while for $k>1$ we obtain $Q(1,1)/Z_k$. For more details about these spaces see \cite{Q11}.} This fixes the bolt location to be given by the
formula: 
\begin{equation*}
\tau _{r}=\frac{kl^{2}+\sqrt{k^{2}l^{4}-900n^{4}+180n^{2}l^{2}}}{30n}
\end{equation*}
and by requiring real values for $\tau_r$ we must restrict the allowed range of the nut charge to be: 
\begin{equation*}
|n|\leq l\frac{\sqrt{90+30\sqrt{k^{2}+9}}}{30}
\end{equation*}

In the $\mathbb{C}$-approach we make the analytic continuations $t\rightarrow iT$ and $n\rightarrow iN$. Then the function $V(\tau )$ is continued to $F(\tau )$ . Imposing the periodicity condition 
\begin{equation*}
\beta _{c}=\frac{4\pi }{|F^{\prime }(\tau _{c})|}=\frac{12\pi |N|}{q}
\end{equation*}%
where $\tau _{c}$ is a root of $F(\tau )$ we find%
\begin{equation*}
\tau _{c}=\frac{ql^{2}+\sqrt{q^{2}l^{4}+900N^{4}+180N^{2}l^{2}}}{30N}
\end{equation*}

It is easy to see now that starting, for instance, with the $\mathbb{C}$-quantities
and using the analytic continuations $N\rightarrow in$ and $q\rightarrow ik$
we recover the corresponding $\mathbb{R}$-quantities and vice-versa. This equivalence
extends to all thermodynamic quantities computed using the counterterm
approach.

\section{Discussion}

The work of this paper was motivated by the observation that the
path-integral formalism can be extended to asymptotically de Sitter
spacetimes to describe quantum correlations between timelike histories,
providing a foundation for gravitational thermodynamics at past/future
infinity \cite{rick2}. The key result is the generalization (\ref{GDuhem})
of the Gibbs-Duhem relation to asymptotically dS spacetimes.

In order to employ this relation it is necessary to analytically continue
the spacetime near past/future infinity. \ There are two apparently distinct
ways of doing this -- the $\mathbb{R}$-approach and the $\mathbb{C}$%
-approach. The $\mathbb{C}$-approach is closest to the more traditional
method of obtaining Euclidian sections for asymptotically flat and AdS
spacetime. The $\mathbb{R}$-approach refers to the Lorentzian section, and
makes use of the path integral formalism only insofar as the generalized
Gibbs-Duhem relation (\ref{GDuhem}) is employed.

The main result of this paper is the demonstration that the $\mathbb{R}$ and 
$\mathbb{C}$-approaches are equivalent, in the sense that we can start from
the $\mathbb{C}$-approach results and derive by consistent analytic
continuations (\textit{i.e.}, using a well-defined prescription for performing the
analytic continuations) all the results from the $\mathbb{R}$-approach.
There are no a-priori obstacles in taking the opposite view, in which the $%
\mathbb{C}$-approach results are derived from the respective $\mathbb{R}$%
-approach results. However, one could still argue that the $\mathbb{C}$%
-approach is the more basic one, as in it the periodicity conditions appear
more naturally than in the $\mathbb{R}$-approach.

On the other hand, the $\mathbb{R}$-approach, when used without the
justification that comes from the $\mathbb{C}$-approach, raises some
interesting questions. Even applied to simple cases such as the
Schwarzschild-dS solution, one may take the view that in the absence of the
nut charge one could still consider a periodicity on the time coordinate in
the Lorentzian sector given by $\beta _{r}=8\pi m$. A more orthodox
interpretation would be that $\beta _{r}$ in the Lorentzian sector is simply
the inverse temperature (as related by the surface gravity of the black hole
horizon) and is not related to a real periodicity of the time coordinate.
Whether or not this is indeed a necessary condition remains to be seen.

Using this equivalence we then proposed an interpretation of the
thermodyamic behaviour of nut-charged spacetimes. In the asymptotically dS
case, we showed that while a subset of the Bolt solutions can have a
sensible physical interpretation, the same does not hold for the Taub-Nut-dS
solutions. Indeed, in the putative Taub-Nut-dS solution the nut is always enclosed in a larger cosmological `bolt' and moreover it does not have a Lorentzian counterpart (\textit{i.e.} it has no equivalent solution in the $\mathbb{R}$-approach). From these facts we conclude that there are no Taub-Nut-dS solutions. This situation holds despite the fact that a naive application of (\ref{GDuhem}) to this case yields thermodynamic quantities that respect the first law of thermodynamics. Rather these quantities are the analytic continuations of their AdS counterparts under $l\rightarrow il$. Similar remarks apply to
the lower-branch dS bolt cases. We have also found that this situation holds in higher dimensions: there are no Taub-Nut-dS solutions, in analogy with the non-existence of the hyperbolic nuts in AdS backgrounds \cite{myers2}.

Moreover, the $\mathbb{C}$-approach has been previously applied with success
to more general cases - it has been proven to be very useful when treating
for instance asymptotically AdS or flat Taub-NUT spaces. In particular, we have shown here that starting from the
well-known results regarding the thermodynamics of the Nut and Bolt solutions in the Euclidian Taub-NUT-AdS case (which corresponds to our $\mathbb{C}$-approach) we can consistently make analytic continuations back to the Lorentzian sections, yielding a physical interpretation of the thermodynamics of such spacetimes. However, this holds only for the bolt solutions;
we found that the Lorentzian AdS-Nut solution did not respect the first law of thermodynamics, rendering the physical interpretation of the Nut solution dubious at best.

We close by commenting on recent results concerning these spacetimes. It has
been shown that there exist broad ranges of parameter space for which
nut-charged spacetimes violate both the maximal mass conjecture and the
N-bound, in both four dimensions and in higher dimensions. However it was
subsequently argued \cite{Balasubramanian:2004wx} that even if the
Taub-NUT-dS spaces do violate the maximal mass conjecture they also
suffer of causal pathologies. However, this is not necessarily the case. As
noted above in both the $\mathbb{R}$ and $\mathbb{C}$-approaches the maximal
mass conjecture can be violated by choosing the parameter $m<0$, independent
of whether or not the metric function has any roots \cite{rick2,rick3}. A
detailed discussion of this situation has appeared recently \cite{Anderson},
where it was emphasized that globally hyperbolic asymptotically dS
spacetimes exist that violate the maximal mass conjecture. In the present
context this will take place whenever the parameters $m$ and $N$ are such
that the function (\ref{VdSTN}) has no roots \cite{rickreview}.

If horizons are present, the maximal mass conjecture and N-bound can both be
violated and we have shown that this holds consistently for both
approaches. Although these cases have regions containing closed timelike
curves (CTC's), our computations pertain to regions where CTCs are absent,
namely outside the cosmological horizon. However, one could argue that such
spacetimes are causally unstable. Whether any spacetimes containing horizons
exist that violate one or both conjectures and that satisfy rigid constraints of
causal stability remains an open question.

In AdS backgrounds it would be interesting to understand how the AdS-CFT correspondence works
in this case, since the Lorentzian section of the Taub-NUT spaces contains closed timelike curves and so it is causally pathological. In fact it was recently noted \cite{Micky1a,Micky1,Micky2,Eugen} that the boundary metric
for the four-dimensional Lorentzian Taub-NUT-AdS spacetime is in fact the
three-dimensional G\"odel metric. This metric also has a bad
reputation as being causally ill-behaved since for generic values of the
parameters, the G\"odel spacetime admits $CTC$'s through every point.
The meaning of a quantum field theory in this background is still an open
problem.

Recently the authors of \cite{GM2} used successfully the $\mathbb{R}$-approach methods to study the Kerr-dS space-times. As the metric is stationary in this case and the continuation to the complex section involves
the analytical continuation of the rotation parameter $a\rightarrow ia$, it
would be interesting to see if there is a similar prescription for the
analytical continuation of the results from the $\mathbb{R}$-approach and
the $\mathbb{C}$-approach and vice-versa. Other interesting extensions of
the present work could involve the study of the electric and magnetic
charged versions of the Taub-NUT solutions (with or without rotation). A new set of nut-charged solutions in Einstein-Maxwell theory has been presented recently in \cite{TNRN,Awad} and it would be interesting to see how one could extend our analysis to these spaces.

\vspace{10pt}
{\Large Acknowledgements}

CS would like to thank Brenda Chng, Tomasz Konopka and Dumitru Astefanesei for valuable discussions during the completion of this work. This work was supported in part by the Natural Sciences \& Engineering Research Council of Canada.

\end{document}